\documentclass[conference]{IEEEtran}
\IEEEoverridecommandlockouts
\usepackage{cite}
\usepackage{amsmath,amssymb,amsfonts}
\usepackage{algorithmic}
\usepackage{graphicx}
\usepackage{textcomp}
\usepackage{xcolor}
\usepackage{makecell}
\usepackage{multirow}
\usepackage{array}
\usepackage{ragged2e}
\usepackage{tabularx}
\usepackage{float}
\usepackage{placeins}
\usepackage{afterpage}
\usepackage{soul}
\usepackage[hyphens]{url}

\def\BibTeX{{\rm B\kern-.05em{\sc i\kern-.025em b}\kern-.08em T\kern-.1667em\lower.7ex\hbox{E}\kern-.125emX}}
\begin{document}

\title{Toward AI Standardization: A Triadic Human-AI Collaboration Framework for Multi-Level Autonomous Mobility\\

\thanks{This work was supported, in part, by the U.S. Department of Transportation (USDOT) University Transportation Center - Mineta Consortium for Equitable, Efficient, and Sustainable Transportation (grant number: 69A3552348328).}
}

\author{%
\IEEEauthorblockN{Gaojian Huang\IEEEauthorrefmark{1}, Wei-Hsiang Lo\IEEEauthorrefmark{1}, Guannan Liu\IEEEauthorrefmark{2}, Yue Luo\IEEEauthorrefmark{1}, and Yantong Jin\IEEEauthorrefmark{1}}
\IEEEauthorblockA{\IEEEauthorrefmark{1}Department of Industrial and Systems Engineering, San Jose State University, San Jose, CA, USA}
\IEEEauthorblockA{\IEEEauthorrefmark{2}Department of Applied Data Science, San Jose State University, San Jose, CA, USA}
\IEEEauthorblockA{Emails: gaojian.huang@sjsu.edu, wei-hsiang.lo@sjsu.edu, guannan.liu@sjsu.edu, yue.luo@sjsu.edu, yantong.jin@sjsu.edu}
}

\maketitle

\begin{abstract}
The goal of the current study is to introduce a triadic human--AI collaboration framework that could be applied in transportation systems such as automated vehicles, micromobility systems, and vehicle teleoperation. Previous standards (e.g., SAE Levels of Automation) have focused on defining automation levels based on who controls the vehicle. However, it is still not clear how human users and AI should collaborate in real-time, especially in dynamic driving contexts where roles can shift frequently. To fill the gap, this study proposed a triadic human-AI collaboration framework with three AI roles (i.e., Advisor, Co-Pilot, and Guardian) that can dynamically adapt to human needs based on real-time data, such as mental states and environmental conditions. The Advisor AI offers informational support without direct intervention; the Co-Pilot AI provides partial intervention when needed, with the goal of sharing control with humans; the Guardian AI performs emergency overrides if necessary. The use cases for these AI roles in the context of micromobility devices (i.e., e-scooters) are presented to demonstrate how these roles can influence user preferences and trust. Overall, the study takes a first step toward a universal role-based collaborative framework for AI standardization and explores how AI technologies can be embedded in future transportation systems while considering human interactions. 
\end{abstract}


\vspace{0.8em}
\noindent\fbox{%
\begin{minipage}{0.96\linewidth}
\small
\textbf{arXiv Version Note.} This manuscript substantially extends and replaces the earlier arXiv preprint titled ``Beyond Levels of Driving Automation: A Triadic Framework of Human-AI Collaboration in On-Road Mobility'' (arXiv:2504.19120v1). The present version reflects the expanded peer-reviewed IEEE CAI 2025 paper, broadening the framework to automated vehicles, micromobility, and vehicle teleoperation. DOI of the IEEE publication: 10.1109/CAI64502.2025.00292.\par\vspace{0.4em}
{\copyright} 2025 IEEE. Personal use of this material is permitted. Permission from IEEE must be obtained for all other uses, in any current or future media, including reprinting/republishing this material for advertising or promotional purposes, creating new collective works, for resale or redistribution to servers or lists, or reuse of any copyrighted component of this work in other works.
\end{minipage}}
\vspace{0.8em}

\section{Introduction}

In recent years, the automotive industry has been rapidly evolving. Many manufacturers and mobility providers now equip vehicles with automated systems, such as Tesla's Full Self-Driving (FSD) system, which allows vehicles to be partially or fully driven by machines instead of human drivers \cite{SAE2021}. With the rapid development of artificial intelligence (AI), these automated systems are becoming more capable of managing driving tasks \cite{yang2024promoting} across diverse environments, ranging from urban to rural areas. In addition, these automated systems extend beyond traditional automated vehicles to include micromobility devices (e.g., e-scooters, e-bikes, and e-wheelchairs), as well as teleoperated vehicles (e.g., robotaxis like Waymo or Tesla Robocabs) \cite{juanico5044936conceptual,tener2023design}. 

With AI systems increasingly integrated into these transportation modes and collaborating with humans in real-time, there is a need to redefine and clarify human--AI collaboration roles. For example, during a semi-autonomous vehicle takeover request prompted by an unexpected construction zone, AI can help drivers decide how to maneuver the vehicle, or provide partial torque on the steering wheel to assist drivers in steering through a shared-control mechanism. Traditional levels of automation classification, such as Driving Automation Levels (SAE International's J3016 standard) \cite{SAE2021} ranging from Level 0 (no automation) to Level 5 (full autonomy), focus on determining who (i.e., human or machine) is in control \cite{Boyle2024,Inagaki2019}. However, these static levels of automation categories do not specify how humans and AI should coordinate in continuous and dynamic driving environments where human--AI interaction patterns shift frequently. In practice, the extent to which automation level is used can vary within a single trip; for example, drivers may rely on Level 3 semi-autonomous functions on the highway and then revert to manual control in dense urban traffic or work zones. Thus, applying a single static classification may fail to accurately represent the dynamic shifts during a trip. 

This mismatch creates practical problems for end-users. Human operators may have difficulties in understanding their own roles as well as the AI's. This means AIs may perform many tasks without keeping drivers in the loop, leading to a lack of system transparency and reduced user trust in the AI; in contrast, AIs may fail to provide adequate support or clear feedback, resulting in user frustration and decreased satisfaction, further lowering user trust~\cite{lee2004trust}. This can be particularly important for vulnerable populations, such as older adults who may be experiencing cognitive and physical declines or people with disabilities \cite{Huang2022a}. Thus, there is a need to standardize AI roles in human-AI interaction and clarify how human drivers should partner with the AI. 

The goal of this study is to propose and demonstrate the application of a novel triadic human-AI collaboration framework. This framework builds upon existing human-automation interaction models~\cite{sheridan2005human,parasuraman2000model} but introduces a novel integration of real-time data, AI system roles, and human input in intelligent mobility systems. It clarifies the roles of the human operator, the AI system, and the data input, which may inform future AI standardization in human-AI interactions.

\section{Related Work}

\subsection{Intelligent Mobility}
Surface transportation has grown rapidly with the advancement of technology. Automated vehicles (AVs) featuring capabilities such as adaptive cruise control, lane-keeping, or conditionally automated driving have become one of the most common vehicles driven on public roads. Based on SAE's Levels of Automation, these vehicles are classified into six levels, from Level 0 (no automation) to Level 5 (fully autonomous) \cite{SAE2021}. One significant distinction between Levels 2 and 3 is the role of human drivers, which shifts from actively monitoring and controlling vehicles (Level 2) to a more passive role in which they only intervene when necessary (Level 3). Today, most AVs on the road are still semi-autonomous and require human drivers to resume manual control at any moment, due to system failures or immediate road events such as missing lane markers or entering a construction zone \cite{McDonald2019}. This transition period, often called a takeover, involves a signal response phase and a post-takeover phase, which includes: 1) perceiving and processing signals for a takeover request, 2) processing information from the driving and vehicle environment to make timely decisions, and 3) executing the maneuvering plan \cite{Huang2022}. 

While more AVs aim to be fully autonomous at Level 5, other applications of AV technology are also being explored. One example is teleoperated driving, where a remote human oversees or partially controls a vehicle when needed. This approach allows remote operators to monitor multiple vehicles simultaneously and handle edge cases, while routine driving tasks are delegated to vehicles' onboard automated systems\cite{tener2023design, kettwich2022helping}. This form of transportation has been used in applications such as robotaxis, autonomous truck fleets, and vehicles operating in hazardous areas. For example, robotaxis can handle most driving tasks themselves; however, when experiencing edge cases, they still need remote human operators to do one of the following: 1) remote driving, where operators directly control the vehicle or share control between the operator and the vehicle, or 2) remote assistance, where operators provide higher-level guidance, such as confirming decisions during uncertain conditions to the vehicle system. Similar to the in-vehicle takeover process, remote operators need time to gain situation awareness and make timely decisions after receiving the remote intervention request. 

Micromobility (e.g., e-scooters, e-bikes, e-wheelchairs) has also benefited from automated systems. These transportation modes, which serve the last-mile needs of short-distance travel (e.g., less than 3 miles)\cite{meroux2023should, gebhardt2022can}, have recently become one of the major travel options in the U.S. \cite{jafarzadehfadaki2024embracing, lee2021forecasting}. Rather than relying on fully autonomous systems, micromobility services typically preserve active rider involvement while receiving targeted system support \cite{european2011white, gerla2014internet}. For example, when users face unexpected scenarios on the road, such as sudden obstacles, this system could send out a warning to alert them about the potential hazard, offer maneuver support (e.g., assisting steering or braking) while the riders perform the action, or directly intervene in the vehicle to handle critical scenarios. Although the role transitions differ across critical contexts, i.e., rider-to-system in micromobility and system-to-human in vehicle takeovers or teleoperation, each of these scenarios requires clear, real-time collaboration between humans and increasingly intelligent automated systems to maximize safety and user experience \cite{fan2022human, rezwana2022understanding}.

With advances in AI, automated systems are becoming more capable, allowing mobility technologies to increasingly assist with, or even manage, the driving task on the road. Such systems represent a form of embodied AI, which can interpret environmental and human inputs (e.g., via advanced sensors and driver biometrics) and deliver real-time, context-aware assistance \cite{de2023autonomous,vanniyakulasingam2023autonomous}. The challenges are now more centered around how to integrate AI in real-time without undermining human oversight or, conversely, overwhelming the user. There is a need for role-based collaboration frameworks that detail ``who does what'' - whether it is a driver in a semi-autonomous car, a remote teleoperator controlling a robotaxi, or a micromobility rider.

\subsection{Human-AI Collaboration Paradigms and Frameworks}

There is a growing emphasis on human-AI teaming in human-AI collaboration research (e.g., \cite{rezwana2023designing}). In this context, AI is considered a co-worker or partner in driving tasks, rather than an assistive tool \cite{berretta2023defining}. As such, humans and AI work together to combine their capabilities to achieve shared goals. This is different from traditional human-automation interaction, where humans may be out of the loop and only intervene when the automation requires a human takeover \cite{parasuraman1997humans}.

Over the years, various frameworks for classifying and implementing human-automation collaboration in vehicles have been proposed, such as supervisory control, shared control, and layered autonomy. Specifically, for supervisory (or traded) control, a human's primary role is to monitor and intervene without continuously acting \cite{sarabia2023review}. This aligns with Sheridan's levels of automation model. Here, one agent, either the human or the AI, is active at a time. In this arrangement, humans delegate tasks to AI and only supervise the system \cite{de2023shared}. This setting applies to SAE Level 2-3 automated driving where human drivers may occasionally intervene and take over.

Shared control indicates that humans and automation simultaneously contribute to the control of the vehicle. For example, the driver holds the steering wheel while an AI co-pilot system applies torque to guide lane-keeping \cite{abbink2012haptic}. In practice, this shared-control paradigm requires both the driver and AI to be actively involved in perceiving the environment and acting on the vehicle, compensating for the other as needed.

As for layered autonomy, the control architecture is mapped into multiple levels, each with different scopes of authority \cite{sarabia2023review}. For example, in teleoperated driving, remote operators typically perform two main roles: higher-level trip planning and lower-level direct control. In the human-AI collaboration paradigm, AI may manage the path and speed (upper layer) while the human operator handles immediate maneuvers (lower layer), or vice versa. The division of responsibilities may depend on factors such as environmental conditions, AI system reliability, operator cognitive workload, and/or specific safety requirements. Another example is Toyota's concept of ``parallel autonomy'' (Toyota Guardian system) \cite{naser2017parallel,Clifford2019}, in which human drivers manage the direction control, while an autonomous safety system continuously monitors the environment and can intervene and override human input when an immediate danger is detected.

While shared control, supervisory control, and layered autonomy provide useful paradigms for allocating responsibilities, these frameworks are limited in their capacity to comprehensively describe human-AI collaboration roles in dynamic and real-time mobility environments. Moreover, they often lack the specificity to distinguish how particular AI roles engage with human users and adapt to different vehicle types, environmental contexts, and task requirements. As AI becomes more embedded in diverse transportation modes, a more flexible and integrated framework is needed to describe how humans and AI should collaborate moment to moment.

\subsection{Existing Automation Taxonomies and AI Standardization}

Current industry standards and guidelines, such as ISO 26262 (Functional Safety for Road Vehicles) \cite{ISO2018} and UL 4600 (Standard for Safety for the Evaluation of Autonomous Products) \cite{ULStandardsEngagement2023}, are mainly concerned with functional safety and hazard analysis. They provide frameworks for mitigating risks such as hardware malfunctions, sensor failures, or software bugs. However, these standards may not specify how an AI system should inform a user in partial automation scenarios or how the AI should escalate from gentle alerts to direct interventions (e.g., forced braking). In other words, the complexities of dynamic human-AI collaboration are not yet fully addressed in these standards and guidelines. 

A similar limitation can be found in SAE International's J3016 standard, which defines six levels of driving automation as described in Section II.A. While J3016 clarifies who is primarily responsible for controlling the vehicle, it does not detail the real-time interplay between humans and AI systems \cite{Boyle2024, Inagaki2019}. For example, AI systems are expected to support drivers during takeover actions in semi-autonomous vehicles (e.g., Levels 2--3), where human drivers need to resume manual control. However, during these transitions, drivers often experience reduced vigilance and situation awareness, highlighting the need for AI to adapt to the driver's real-time mental state and surrounding conditions. A current challenge to smooth transitions is the lack of clarity about roles between human drivers and AI, making it uncertain which specific tasks each party should handle. This could be especially problematic in scenarios where automation levels shift within a single routine trip, such as switching from conditional automation (i.e., SAE Level 3) on highways to manual control in urban areas. In contrast to sudden takeovers, these transitions driven by changing driving contexts (e.g., switching from highway automation to urban manual control) require proactive coordination between human and AI systems. This ambiguity can lead to critical issues such as limited understanding of the automation state, miscalibrated trust (e.g., over- or under-reliance), and inconsistent or sub-optimal decision-making.

While more emerging AI standards are currently under development, there is a need to move toward a human-centered AI approach\cite{shneiderman2022human} that includes explicit frameworks that go beyond static ``levels of control'' to define evolving roles within continuous human-AI interactions.

\begin{table*}[htbp]
    \centering
    \caption{Triadic AI Roles in Human-AI Collaboration: Control Scopes, Key Functions, and Adaptive Behaviors Across Mobility Domains}
    \begin{tabularx}{\textwidth}{|c|>{\centering\arraybackslash}X|>{\centering\arraybackslash}X|>{\centering\arraybackslash}X|}
    \hline
    \textbf{Dimension} & \textbf{Advisor} & \textbf{Co-Pilot} & \textbf{Guardian} \\
    \hline
    \textbf{Control Scopes} & 
     \makecell[l]{- No direct vehicle control \\ 
     - Monitors driving conditions \\
     - Offers purely informational alerts, \\ prompts, and suggestions \\
     - Escalates alerts if the driver is \\inattentive or if hazards become critical} & \makecell[l]{- Shared/partial control \\ 
     - Dynamically adjusts steering, braking, \\ or speed when necessary \\
     - Allows driver override at any time \\
     - Provides active assistance \\(steering nudges, speed modulation) \\to maintain stability and safety}
     & \makecell[l]{- Performs emergency override \\ 
     - Assumes full control \\ in life-threatening situations \\
     - Overrides erroneous driver \\ input if it poses severe danger \\
     - Typically operates quietly in the \\ background} \\
    \hline
    \textbf{Key Functions} & 
    \makecell[l]{- Informational alerts (hazards, route \\ changes, environmental conditions) \\
     - Context-aware guidance (traffic, weather, \\upcoming maneuvers) \\
     - Driver state monitoring (physiological \\ monitoring) for soft reminders (fatigue, \\ break suggestions)} & 
     \makecell[l]{- Maneuver assistance (lane changes, \\merges, brakes) \\ 
     - Partial automation (lane-centering, \\ adaptive cruise, speed assist) \\
     - Driver state monitoring (eye tracking, \\ grip strength) for real-time assistance}
     & \makecell[l]{- Real-time hazard monitoring (e.g., \\ collision risk $>$ $95\%$) \\ 
     - Immediate intervention (evasive steering,\\ emergency braking) \\
     - Stabilization (correct skids, hydroplaning)} \\
     \hline
    \makecell[l]{\textbf{Adaptive} \\ \textbf{Features}} & 
    \makecell[l]{- Adjusts alert urgency based on driver \\ state (distraction) or environment severity \\
     - Scales back non-critical notifications to \\ avoid alert fatigue} & 
     \makecell[l]{- Escalates assistance if the driver is \\inattentive, the environment is complex\\(dense traffic, inclement weather),  \\or user input is insufficient\\
     - Reduces involvement when the driver \\re-engages, passing control back smoothly}
     & \makecell[l]{- Performs as a continuous safety governor \\ 
     - Provides post-intervention transparency\\by explaining actions afterward \\
     - Deactivates once the critical event has \\ passed} \\
    \hline
    \makecell[l]{\textbf{Examples} \\ \textbf{(Automotive)}} & 
    \makecell[l]{- ``Let's merge left."\\
     - ``Sharp curve in $600$ feet''\\
     - ``We've been driving two hours; \\ want to find a rest stop?''} & 
     \makecell[l]{- ``I'll apply slight steering torque.'' \\
     - ``You seem distracted; I can handle speed \\ until you're ready.''\\
     - Automatically slows if the driver misses \\ a stop sign.}
     & \makecell[l]{- ``Emergency braking: \\ collision avoided - you're safe now.'' \\
     - ``Steering correction applied: \\ roads are icy.''\\
     - (Post-crisis) ``I took control because a \\ collision was imminent. Are you alright?''} \\
    \hline
    \makecell[l]{\textbf{Examples} \\ \textbf{(Micromobility:} \\\textbf{e-Scooters)}} & 
    \makecell[l]{- ``Watch out for pedestrians at next \\ crosswalk.'' \\
     - ``Battery low in $2$ miles - consider a \\ charging stop.''} & 
     \makecell[l]{- ``I'll help stabilize steering in these \\ narrow lanes - shall I?'' \\
     - ``I'll limit speed \\ slightly to ensure control.''}
     & \makecell[l]{- Automatically engages the e-scooter's \\ emergency brake if a sudden obstacle \\ appears and the rider fails to react. \\ - To riders: ``Braked to prevent a collision.''} \\
     \hline
    \makecell[l]{\textbf{Examples} \\ \textbf{(Teleoperation)}} & 
    \makecell[l]{- In a remote driving center, displays \\ real-time sensor data to the teleoperator: \\ ``Vehicle behind you is approaching fast; \\ consider a lane change soon.''} & 
     \makecell[l]{- Offers partial steering or speed control \\ if the remote driver is momentarily \\overwhelmed. ``I'm providing a gentle \\ steering assistance - confirm to proceed.'' }
     & \makecell[l]{- If the teleoperator is unresponsive and \\ a crash is imminent, it triggers an \\ emergency maneuver (e.g., remote override \\ of throttle/brakes). \\ - (Post-event) ``Override executed due to \\ collision risk - vehicle is stable.''} \\
    \hline
    \makecell[l]{\textbf{Companion/} \\ \textbf{Emotional} \\ 
    \textbf{Support}} & 
    \makecell[l]{- Conversational tone (``How are you \\ feeling?'') \\
    - Encourages breaks if fatigue is detected \\
    - Reduces isolation: chatty voice-based \\ interaction if driver prefers} & 
     \makecell[l]{- Empathetic prompts (``I'll handle the \\ next few minutes while you calm down.'') \\
     - Lighthearted banter or reassurance to \\ de-stress driver \\
     - Encourages safer decisions (seatbelt \\ checks, speed moderation)}
     & \makecell[l]{- Typically less conversational \\ - Post-event reassurance (``You're safe \\ now; do you want to pull over and rest?'') \\
     - Explains why override happened \\ (reduces confusion or distrust)} \\
    \hline
    \makecell[l]{\textbf{Typical Use} \\ \textbf{Cases (SAE} \\ 
    \textbf{Levels)}} & 
    \makecell[l]{- Level 0-1: Basic alerts (seatbelt \\ reminders, hazard beeps) \\
    - Level 2-3: More robust situation \\ prompts (traffic, route info, takeover \\ requests) \\
    - Level 4-5: Emphasizes user preferences \\ \& comfort info} & 
     \makecell[l]{- Level 0-3: Lane-centering, adaptive \\ cruise, partial maneuvers during driver \\ inattention \\
     - Level 4-5: Optional co-control based on \\user preferences (e.g., scenic routing, \\speed adjustment, or allowing the user to \\``practice'' driving)}
     & \makecell[l]{- All levels (where applicable): Guardian \\ might only intervene when a collision\\
     is imminent \\
     - Level 4-5: Full override in rare system \\ failures (e.g., sensor malfunction, \\ user incapacitation)} \\
    \hline
    \end{tabularx}
    \label{tab:AI_roles}
\end{table*}

\section{The Triadic Framework for Human--AI Collaboration}
\label{section3}
To fill the identified gap, this study developed a triadic framework of AI roles that can switch fluidly in real time, aligned with the dynamic and adaptive nature of human-AI collaboration in complex transportation environments. We conducted a broad literature review of frameworks and role classifications in human-centered AI, human-AI teaming, human-automation interaction, and human-centered embodied AI across a wide range of application domains, as well as classical human factors models such as Sheridan's Levels of Automation and Endsley's Situation Awareness framework \cite{sheridan1978human,endsley1995toward}. From the review, we classified key functions that AI may offer in a real-time driving environment and proposed a triadic human-AI collaboration framework with three distinct and non-hierarchical AI roles in terms of collaboration boundaries and communication strategies: Advisor, Co-Pilot, and Guardian AI.

\paragraph{Advisor AI} This AI role provides the human user (driver, rider, or remote operator) with continuous informational support, such as context-aware alerts or informational suggestions, without directly intervening in vehicle control. This AI role can prevent driver complacency or confusion by providing timely updates, keeping the drivers informed and engaged.
\paragraph{Co-Pilot AI} Co-Pilot AI shares partial control with the human user by offering dynamic action support that adapts to human action, such as adjusting steering, braking, or speed when user input is insufficient, attention is compromised, or external conditions become complex. This AI role reduces driver workload without removing the driver from the control loop. 
\paragraph{Guardian AI} This AI role serves as a continuous safety governor, autonomously taking full control only to prevent collisions or perform emergency maneuvers. It can even override driver inputs when erroneous actions and/or severe danger are predicted. 

These three AI roles are not locked to a fixed setting; instead, they transition dynamically according to user state and situational demands. For example, if repeated alerts from Advisor AI are ignored, or if the system detects driver distraction/fatigue, it may shift to Co-Pilot AI. Similarly, in scenarios where partial assistance cannot adequately prioritize safety, Guardian AI can step in with a full override. Guardian AI can revert to either Advisor or Co-Pilot mode as well, based on user needs and environmental context. For instance, experienced drivers may prefer to remain highly engaged, or e-scooter riders may prefer manual riding for entertainment purposes \cite{wang2023travel,christoforou2021using}.
Table~\ref{tab:AI_roles} consolidates and extends the earlier preprint version of this framework by detailing the three AI roles (Advisor, Co-Pilot, and Guardian) in human-AI collaboration, including their control scope, key functions, adaptive features, and practical applications across diverse mobility domains including automated vehicles, micromobility devices, and teleoperation scenarios.

\section{Integrating Real-Time Data from Humans and the Environment}
\begin{figure*}[!t] 
    \centering \includegraphics[width=2\columnwidth]{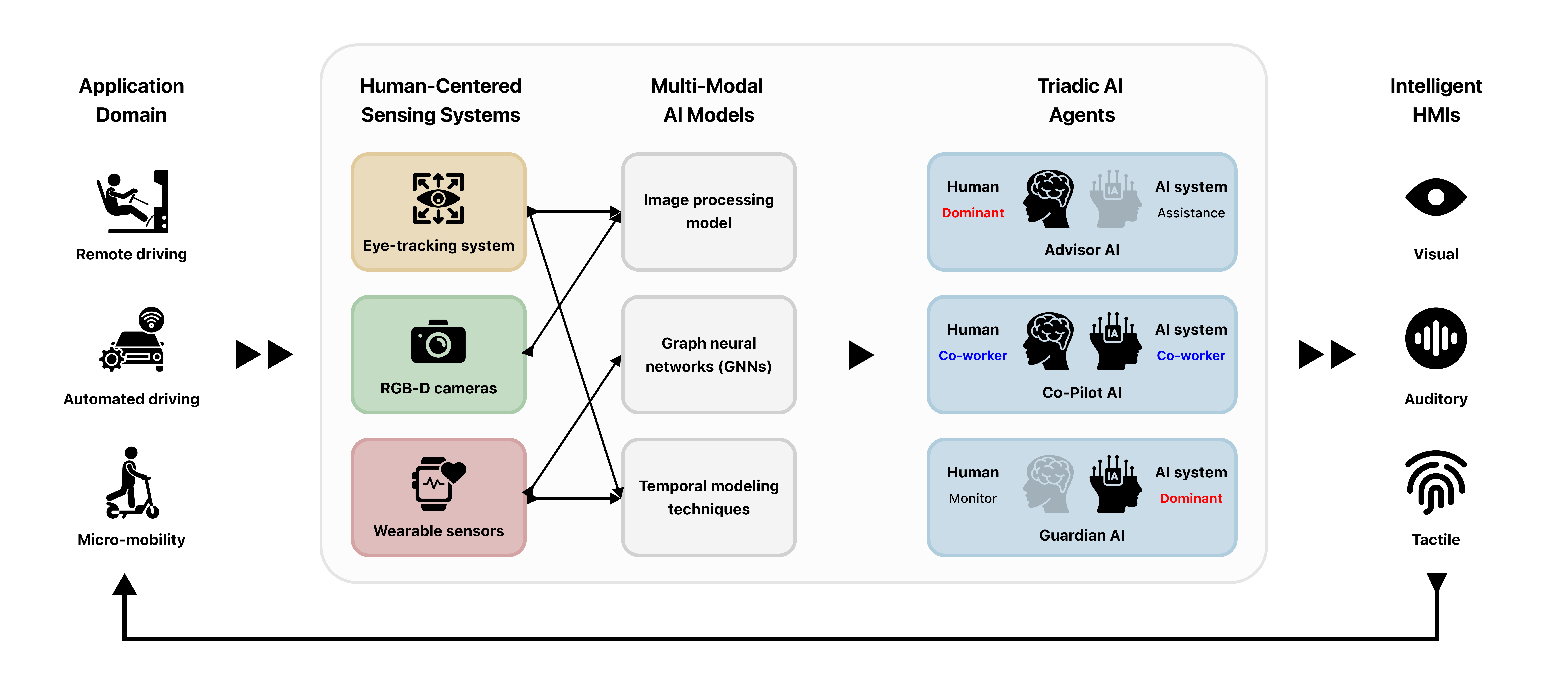}  \caption{System architecture illustrating the scenario of a human interacting with an intelligent mobility system.}
    \label{fig:diagram} 
\end{figure*}
Fig.~\ref{fig:diagram} illustrated the system architecture of our proposed triadic framework, highlighting how real-time human and environmental data is used to inform adaptive role transitions among AI agents for human-AI collaboration. As illustrated in Fig.~\ref{fig:diagram}, the framework begins with human-centered sensing systems, which include cameras (e.g., RGB-D cameras) and sensors (e.g., eye tracking devices), to continuously acquire multi-modal data streams from both the user and the surrounding environment. These data streams are then processed by specialized multi-modal AI models. For example, image processing models are applied to visual data captured from eye-tracking systems and RGB-D cameras, enabling human gaze analysis and environmental perception. Graph neural networks are employed to model skeletal structures derived from wearable sensor data, facilitating the interpretation of human postures and motion dynamics. Furthermore, advanced temporal modeling techniques are utilized to process high-dimensional time-series data in order to capture underlying physiological patterns and temporal dependencies. Building upon these multi-modal AI models, the triadic framework is implemented via AI agents, each representing a distinct collaboration role: the Advisor AI, the Co-Pilot AI, and the Guardian AI, which interpret and respond to real-time user and environment context. Through this three-role structure, we may further design intelligent human-machine interfaces incorporating multiple sensory channels (e.g., visual, auditory, and tactile modalities) to convey information from AI agents. These interfaces facilitate intuitive and adaptive interactions between the user and each AI role, particularly within complex and dynamic transportation environments. The details are described as follows.

\subsection{Human-Centered Sensing Systems}
\label{sec:data_acquisition}

The transition of AI's role, as discussed in Session \ref{section3}, depends on the user's state, which can be inferred through multiple layers of human physical and physiological data, each requiring different types of sensors for accurate detection. 
At a fundamental level, macro-level indicators such as posture and gestures provide interpretable human behavioral data \cite{mahomed2024driver,espericueta2025using}. These can be captured using vision-based devices, including RGB cameras, depth sensors, and wearable motion trackers, which analyze body movements and seating positions \cite{meda2025insights} to infer users' physical intention and engagement.
On a more subtle level, gaze patterns and facial expressions are associated with users' cognitive states \cite{prasse2024improving,lohani2019review}. These cues, though less overt than body movements, can be detected using eye-tracking and other vision-based devices \cite{luo2023multisensory}. These sensors monitor human attention expressions, determining the user's level of focus and potential distraction.
At a deeper physiological level, ``under-the-skin'' signals such as heart rate variability, skin conductance, and other biometric markers offer assessments of mental states. These indicators can be measured using wearable sensors (e.g., smartwatches, chest straps, or capacitive seat sensors) to detect stress, fatigue, drowsiness, heightened alertness, and other mental states \cite{rimes2017stress,smith2020using}.

In this triadic framework, the AI roles can shift based on user state and environment context. Similarly, human's operational role, whether as a driver, rider, or remote operator, influences their expectations of AI involvement and the type of data needed for effective collaboration.

For drivers, whether assisted by automation or not, humans are still responsible for vehicle control according to current regulations. AI, in this case, can cycle among all three AI roles depending on need. It draws on macro-level behavioral data (e.g., gestures for commands \cite{mahomed2024driver}), micro-level cognitive data (e.g., gaze, blink rate for focus \cite{gavas2020blink}), and physiological data (e.g., heart rate variability for fatigue detection \cite{lu2022detecting}) to tailor feedback, issue timely warnings, share partial steering control, or execute a full override of braking. This multi-role adaptability provides safe and effective driving support.

For riders, who delegate driving responsibilities to either an AI system or another human, AI's role centers on safety and comfort. To provide a smooth and reassuring ride, AI needs to actively communicate with riders with status updates, helping them maintain situation awareness despite not being in control. The AI should dynamically adjust the vehicle's behavior based on riders' preferences. This is achieved by leveraging engagement and preference data, such as facial expressions, physiological signals (e.g., heart rate, skin conductance), and verbal and non-verbal cues \cite{hu2024exploring}. Through continuous interpretation of these signals, the AI can tailor its actions and interactions to foster trust and enhance the overall riding experience.

For remote operators who may oversee multiple AI-driven vehicles, AI plays a key role in managing information flow and decision support. Unlike drivers, remote operators rely on real-time data streams of vehicle telemetry (e.g., speed, system status) and sensor feeds (e.g., camera, LiDAR data) from the environments, as well as network metrics such as latency or bandwidth that affect command timing \cite{ortiz2020vehicle}. Using behavioral (e.g., body orientation for vehicle focus \cite{raza2018appearance}), cognitive (e.g., pupil dilation for cognitive load \cite{zheng2020comparison}), and physiological indicators (e.g., galvanic skin response for alertness \cite{nawawi2023drowsiness}), the AI filters critical alerts, prioritizes tasks, and suggests or executes interventions for synchronizing both human and system actions.
\begin{figure*}[!t] 
    \centering \includegraphics[width=2\columnwidth]{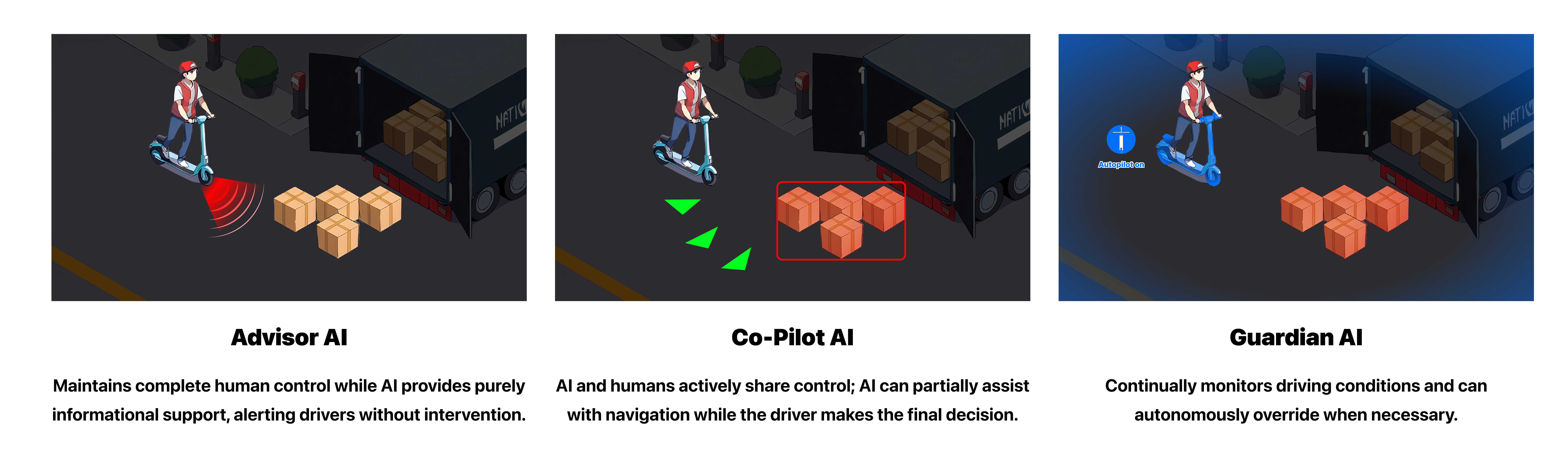}     \caption{Application of three AI roles in critical micromobility scenarios.}
    \label{fig:AI role} 
\end{figure*}

\subsection{Multi-Modal AI for Heterogeneous Data Processing}
Human-AI collaboration can be further enhanced by an integrated framework that acquires and fuses heterogeneous data streams from both human-centered sensing and environmental sensing systems. While Section~\ref{sec:data_acquisition} focused on human-centered data (e.g., physiological signals), environmental sensors such as radar and LiDAR provide complementary information that further enhances the AI's understanding of the surrounding content. Radar can be used to detect object velocities and distances, providing crucial information about moving obstacles~\cite{10571852, 10311072, 9465646}. LiDAR, on the other hand, offers high-resolution three-dimensional maps of the surrounding environment for precise detection of objects, road boundaries, and free space~\cite{10311072}. Combining these environmental sensors with human-centered data allows for a comprehensive understanding of both the user and the user's surrounding environment.

\raggedbottom

To effectively process the diverse data formats collected from human-centered and environmental sensors, several advanced AI models can be leveraged. For image data, cutting-edge models such as Vision Transformers (ViTs)~\cite{dosovitskiy2020image}, Convolutional Neural Network (CNN)~\cite{726791}, and ResNet~\cite{he2016deep} are used to capture spatial and semantic features of the environment. In parallel, depth sensor data representing human skeletal motion are transformed into graph structures, where joints are represented as nodes, and their anatomical connections as edges. Graph neural networks (GNNs) are employed to capture these spatial and temporal relationships that include complex human postures, gestures, and movement sequences~\cite{10706704, 9828603}. Continuous time-series signals collected from wearable sensors require sophisticated signal processing approaches. Temporal modeling techniques, including Transformer~\cite{vaswani2017attention}, temporal convolutional networks (TCNs)~\cite{lea2017temporal}, and advanced time-series decomposition approaches~\cite{10752880, 10752411}, can be applied to capture the underlying physiological rhythms and anomalies. Furthermore, radar and LiDAR data can be processed effectively using models such as PointNet~\cite{qi2017pointnet} or VoxelNet~\cite{zhou2018voxelnet}, which are well-suited for handling three-dimensional point clouds. Collectively, these multi-modal AI models support the robust integration of human-centered and environmental data and establish the technical foundation for the triadic AI roles.

\subsection{Triadic Framework for Real-Time Collaboration}
Building on the foundational multi-modal AI models, we implement the triadic framework to function in complex transportation environments. As stated in Section~\ref{section3}, these role-specific AI agents vary in autonomy and user engagement yet draw on the same integrated data streams to determine when and how to act.

The Advisor AI agent uses the underlying AI models to provide data-driven insights, alerts, and contextual observations. This agent serves as an analytical companion to provide timely information and suggestions while leaving decision-making authority and final control with the user.

The Co-Pilot AI agent represents a shared-control partner that assists in decision-making and low-level actuation. The Co-Pilot AI continuously engages with the user for optimal actions, highlighting potential risks and adapting steering, braking, or speed in real-time based on both environmental changes and user feedback. This layer provides a human-AI partnership in which the agent reduces human cognitive and physical workload without removing the human from the control loop.

The Guardian AI agent acts as a safety governor, guided by predictive modeling and learned behavioral patterns. When multi-modal data indicate an immediate hazard, the Guardian AI agent actively initiates or directs actions on behalf of the user, with minimal need for human input. 

The integration of multi-modal data into the triadic AI framework enables various levels of support, from passive advisory functions to fully autonomous guidance. Furthermore, given that each role may change in real time, it is essential to design intelligent human-machine interfaces (HMIs) that effectively communicate these transitions so that users can understand AI behavior and enhance appropriate levels of trust and user acceptance.

\begin{figure*}[!t] 
    \centering \includegraphics[width=2\columnwidth]{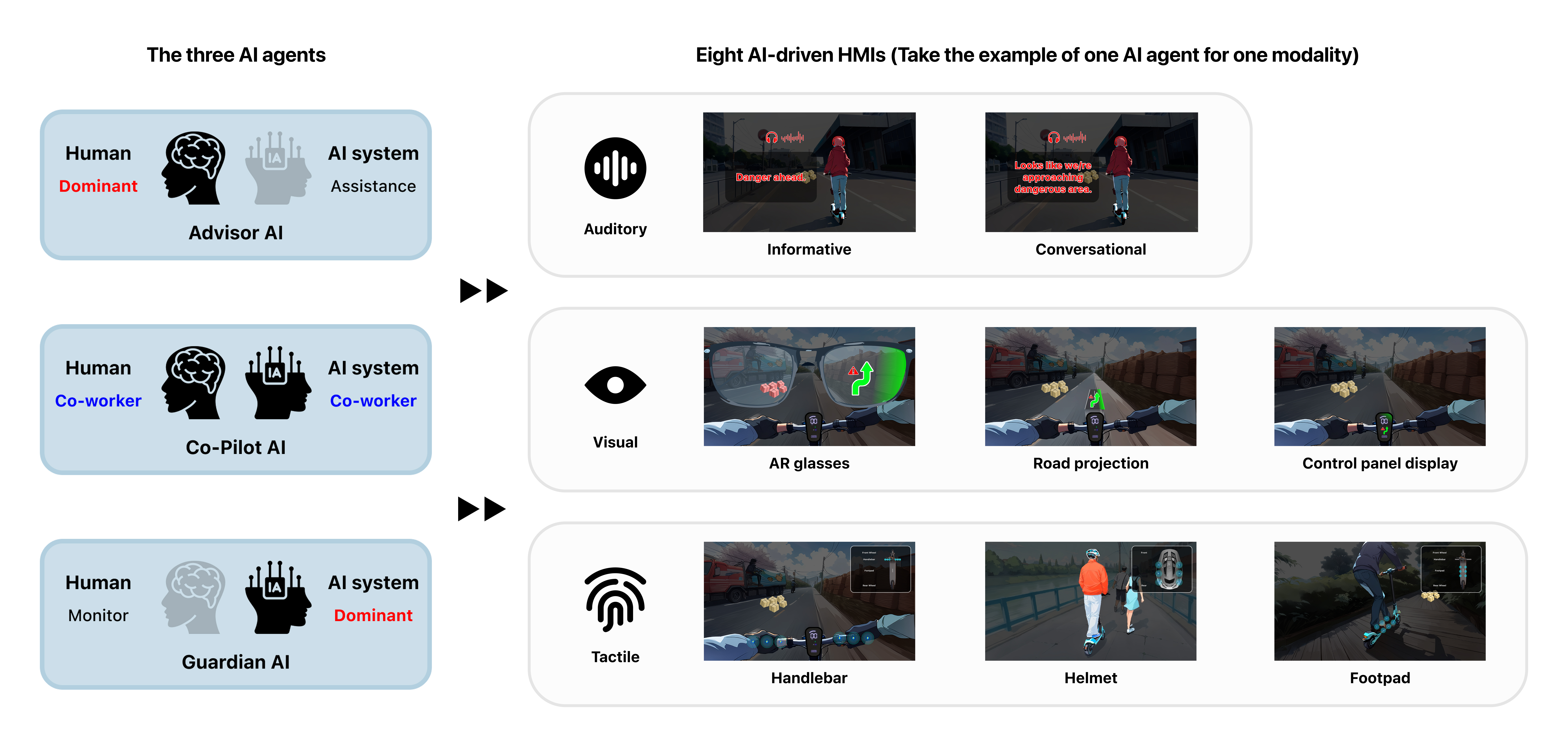} 
    \caption{The example of the triadic AI framework in eight HMIs (e.g., Advisor AI in the auditory format, Co-Pilot AI in the visual format, and Guardian AI in the tactile format).}
    \label{fig:HMI} 
\end{figure*}

\section{Application Scenarios and Case Studies}
In this section, we present an online study focusing on micromobility (i.e., e-scooters) as a preliminary validation of the Triadic Human-AI Collaboration Framework\cite{Lo2025}. While integrating AI-driven systems is a potential solution \cite{european2011white, gerla2014internet} to mitigate safety issues caused by the increasing usage of e-scooters \cite{ma2021scooter}, the precise role of AI in micromobility has not yet been systematically investigated. To address this gap, we conducted an online survey \cite{Lo2025} to compare the effect of three AI roles (i.e., Advisor, Co-Pilot, and Guardian, Fig.~\ref{fig:AI role}) on user preference. An online approach allowed a large and demographically varied sample to be obtained in a relatively short period, which can provide early feasibility data before subsequent in-lab studies for objective evaluation. 

The three AI agents were presented through human-machine interfaces (HMIs) in three modalities, visual, auditory, and tactile, similar to previous driving studies (e.g., \cite{bazilinskyy2018take, Huang2022}). To increase the variations of the HMIs, each visual, auditory, and tactile HMI was further developed into multiple types (Fig.~\ref{fig:HMI}), including three visual types: AR glasses, control panel display, and road projection; two auditory types: informative (e.g., ``Exit ahead'') and conversational (e.g., ``We are entering a new road.'') voice assistance; and three tactile types: handlebar, footpad, and helmet.

The survey was conducted on a crowdsourcing platform, Prolific, to compare user preferences for the three AI agents and three modalities (nine role-by-modality combinations, see Fig.~\ref{fig:HMI}) illustrated with eight HMI types. A total of $473$ valid responses (mean age = $46.29$) were collected. To systematically evaluate user preference, the questionnaire, which included usefulness and satisfaction ratings, was adopted \cite{van1997simple}. The results revealed that no statistically significant differences were found among the three AI roles, regardless of usefulness or satisfaction ratings. In terms of the HMI modality, auditory stimuli offered greater usefulness and satisfaction compared to visual and tactile options. Within the three visual HMIs, the control panel was the least useful compared to AR glasses and road projection. However, regarding satisfaction, AR glasses were the least preferred. The findings for auditory HMIs demonstrate that informative assistance was more useful and satisfying than conversational assistance. Among the comparisons of the three tactile HMIs, results suggest that incorporating displays on handlebars was associated with higher usefulness and satisfaction than footpads or helmets.

Overall, the findings of this study provide early evaluations of the Triadic Human-AI Collaboration Framework for user preferences, which could inform the design of follow-up studies on 1) user preferences for other transportation modes, such as automated vehicles or teleoperated vehicles, and 2) objective validation of the framework in controlled experiments or real-world trials; and (3) development of regulatory and standards frameworks that address human-AI collaboration across vehicle types, including personal, shared, and commercial mobility systems.

\section{Discussion and Future Work}
\subsection{Conceptual Contributions}
This study proposes a triadic human-AI collaboration framework (Advisor, Co-Pilot, and Guardian) to fill gaps in the literature and specify how responsibility shifts among these AI roles across transportation domains. The level of AI involvement should be determined by real-time sensing of both the human state and environmental conditions. One core contribution of this study is to define AI roles that map directly onto the diverse tasks required in intelligent mobility. While earlier frameworks (e.g., SAE Levels of Automation) emphasize who is in control, this triadic approach clarifies how control may flow among distinct AI roles, reducing role ambiguity between AI and drivers, riders, or remote operators.

\subsection{Implications for Industry and Standards}
The role-based framework proposed in this study has important implications for ongoing and future AI standardization efforts. For example, industry guidelines (e.g., ISO 26262 and UL 4600, or new AI guidelines that become available in the near future) could incorporate ``AI role protocols'' to specify the boundaries for triggering the movement of the system from Advisor to Co-Pilot mode, or how Guardian overrides need to be documented and communicated to the users. Standardized role definitions may facilitate audits, certification processes, and legal or liability considerations. Explicit definitions of how AI roles respond to real-time sensing can also facilitate interoperability across automakers, micromobility platforms, and teleoperated services while helping users consistently interpret system actions and safety overrides. Additionally, the framework highlights the need to address potential ethical and legal challenges, such as determining liability in cases where a Guardian AI override leads to unintended consequences, as well as ensuring interoperability with existing regulatory and technical standards.

\subsection{Communication Strategies: What/Action-Focused (Directive) vs. Why-Focused (Explanatory)}
An additional dimension that may complement the triadic framework proposed in this study is the AI communication strategy, that is, how AI presents information to humans, either through direct alerts or commands where AI tells the human what to do (or what the AI is doing) without additional context. For example, the car may issue a sharp warning: ``Brake now!'' when it detects an immediate hazard. On the other hand, explanatory AI may prioritize communicating ``why'' through a brief justification or context for the request, such as ``Take over, construction zone ahead.'' Direct alerts have the advantage of being quick and unambiguous, while explanatory communication may increase the transparency of AI systems. How the two communication strategies interact with the three AI roles under different circumstances (e.g., mild hazard vs. immediate threat) requires further research.

\subsection{Adaptive HMIs as the Interface for AI Roles}
Because each of the three AI roles involves different levels of interaction, adaptive human-machine interfaces (HMIs) play a critical role in conveying which AI agent is active at any given moment. For example, Advisor AI may rely on minimal auditory alerts, while Co-Pilot might highlight shared steering controls in the user interface (UI). Similarly, with different AI communication strategies, either What/Action-Focused (Directive) or Why-Focused (Explanatory), the modality used to convey HMIs may vary. The examples shown in Table I are in a conversational format, which primarily relies on the auditory channel. In real-life time-critical events, users may need multimodal HMIs (i.e., combined visual, auditory, and/or tactile feedback) to quickly perceive information from AI. Finally, for people with different perceptual or cognitive abilities, HMIs should also be adaptive to effectively communicate with human users. For example, older adults with general sensory decline may experience fluctuating day-to-day cognitive functioning. In this case, for effective interaction, AI may need an adaptive HMI that adjust its parameters (e.g., intensity and duration) and modalities in response to users' real-time states and environmental conditions.

\subsection{Broader Application and Empirical Validation}
Although this study primarily addresses automated vehicles, micromobility, and vehicle teleoperation, the triadic framework may be extended to other high-stakes domains. For example, healthcare robotics could use Advisor AI to suggest adjustments or flag potential issues, Co-Pilot AI to share partial control with a medical practitioner, and Guardian AI to intervene when a critical error is detected. Further work is needed to test the framework's feasibility across these applications. Even within the surface transportation domain, although conceptually robust, the triadic framework has yet to be validated for real-world effectiveness through simulations, in-lab controlled experiments, or field studies, particularly in relation to critical factors such as safety, trust, and operators' workload. Increasing the complexity of testing conditions may yield more realistic and generalizable insights, especially when combined with measurable indicators such as reaction time, level of human intervention, or physiological responses in human-AI teaming scenarios. Building on this foundation, future research may examine the effects of the three AI roles on driving, riding, or operating performance; trust in automation and AI; perceived usefulness and satisfaction across diverse participant groups, including novices, expert drivers, older adults, or individuals with disabilities.

\section{Conclusion}
This study introduces a Triadic Human-AI Collaboration Framework, featuring three distinct AI roles---Advisor, Co-Pilot, and Guardian, to enhance AI standardization and real-time cooperation in transportation systems. The framework establishes a dynamic approach to AI involvement by adapting to human states and environmental conditions. Further empirical validation, through controlled experiments and real-world trials, is necessary to refine its applicability across diverse domains. Overall, this study lays the groundwork for both future AI standardization and human-AI collaboration in AI-driven transportation systems.

\bibliographystyle{IEEEtran}
\bibliography{references}

\end{document}